\documentstyle[draft]{jaa}
\input epsf.sty
\begin{document}

%%%%%%%%%%%%%%%%%%%%%%%% Title Page begins %%%%%%%%%%%%%%%%%%%%%%%%%%%%%

\title[FIR mapping of W3(OH), S 209 \& S 187 regions]
{Far infrared mapping of three Galactic star forming regions :
 W3(OH), S 209 \& S 187 }

\author[S.K. Ghosh et al]
{S.K. Ghosh$^{1,2}$\thanks{E-mail:swarna@tifr.res.in},
B. Mookerjea$^{1,3}$, 
T.N. Rengarajan$^4$,
\newauthor
S.N. Tandon$^5$ and R.P. Verma$^1$\\
$^1$Tata Institute of Fundamental Research, 
Bombay 400 005, India \\
$^2$Institute of Space and Astronautical Science, Kanagawa 229, Japan\\
$^3$JAP, Indian Institute of Science, 
Bangalore 560 012, India\\
$^4$Department of Physics, Nagoya University, 
 Nagoya 464-8602, Japan\\
$^5$Inter-University Centre for Astronomy 
\& Astrophysics, Pune 411 007, India}

\maketitle
%\vskip -2.5cm
 Running title : FIR mapping of W3(OH), S 209 \& S 187 regions
%\vskip 1.5cm
%\begin{verbatim}
%          Address for communication :
%                     S. K. Ghosh
%                     Infrared Astronomy Group (DAA)
%                     Tata Institute of Fundamental Research, 
%                     Homi Bhabha Road, Colaba
%                     Mumbai (Bombay) 400 005 
%                     India
% 
%\end{verbatim}
%%%%%%%%%%%%%%%%%%%%%%%%%%% Title Page ends %%%%%%%%%%%%%%%%%%%%%%%%%%%%%
%%%%%%%%%%%%%%%%%%%%%%%%%% Abstract begins %%%%%%%%%%%%%%%%%%%%%%%%%%%%%
%\newpage
\begin{abstract}

   Three Galactic star forming regions 
associated with W3(OH), S209 and S187 have been simultaneously
mapped in two trans-IRAS far infared (FIR) bands centered at
$\sim$ 140 and 200 $\mu$m using the TIFR 100 cm balloon borne FIR telescope. 
These maps show extended FIR emission with structures.
The HIRES processed IRAS 
maps of these regions at 12, 25, 60 \& 100 $\mu$m have also been presented 
for comparison. 
Point-like sources have been extracted from the longest
waveband TIFR maps and searched for associations in the other five bands.
The diffuse emission from these regions have been quantified,
which turns out to be a 
significant fraction of the total emission.
The spatial distribution of cold dust (T $<$ 30 K) for two of these 
sources (W3(OH) \& S209), has been determined reliably from the maps in
TIFR bands. 
The dust temperature and optical depth maps show complex 
morphology. In general the dust around S 209 has been
found to be warmer than that in W3(OH) region.

\end{abstract}


%%%%%%%%%%%%%%%%%%%%%%%%%% Abstract ends %%%%%%%%%%%%%%%%%%%%%%%%%%%%%%%

%%%%%%%%%%%%%%%%%%%%%%%% Main Text begins %%%%%%%%%%%%%%%%%%%%%%%%%%%%%%
\newpage
  
\section{Introduction}

 The far infrared (FIR) continuum emission from the interstellar
dust component allows one to probe deeper in to the denser
regions of Galactic star forming regions.
A long term programme of studying Galactic star forming regions
is being pursued at the Tata Institute of Fundamental Research (TIFR) 
using its 100 cm
balloon borne FIR telescope. This programme 
aims at high resolution ($\sim 1^{\prime}$) mapping in two FIR bands centered 
at wavelengths $\sim$ 150 and 200 $\mu$m, 
beyond the longest waveband of IRAS survey.
The trans-IRAS wavebands help in detecting colder component
of the dust.
Several Galactic star forming regions have been studied 
leading to detection of cold dust (upto 15 K) and its 
spatial distribution (e.g. Ghosh et al 2000, Mookerjea et al 2000).
 The present study deals with three regions selected on
the basis of 
their association with powerful molecular outflow activity and
their extended / complex morphology. These are : W3(OH), S 209 and S 187
regions.

 The Galactic star forming region known as W3(OH),
is a very unique and interesting source for several reasons.
It is situated $\sim 13^{\prime}$ SE of W3 (main) in the giant molecular cloud,
along the prominent ridge of star formation in the Perseus arm
at a distance of 2.3 kpc.
It is one of the most luminous high emission measure compact H II
region of shell type morphology (Dreher and Welch 1981).
Surrounding the ionized gas, there exist dense molecular
clumps which host spectacular sources of OH, H$_{2}$O and CH$_3$OH
maser emission as well as a bipolar outflow source (Wink et al 1994).
Tieftrunk et al (1998) have surveyed this region in
the NH$_3$ line in which they have detected extended emission.
A strong far infrared source is associated with the H II region
(Campbell et al 1989).
W3(OH) has received a lot of attention recently
from cm, mm and sub-mm
waveband researchers, though mostly
concentrating on the higher spatial resolution of the
very central few arc sec region.
Here we present the study of the distribution of dust 
in the general neighbourhood of W3(OH) (within a few parsec).

The S209 region is an evolved H II region with visible optical
nebulosity in the outer Galaxy. 
The ionized region is very extended and luminous in radio continuum.
 The emission at 1.4 GHz has been detected over 
12${}^{\prime} \times$7${}^{\prime}$
by Fich (1993). The associated molecular gas extends over 
a region of $\sim 14^{\prime}$ diameter,
as inferred from the CO survey of Blitz, Fich \& Stark (1982).
The CO line velocity places S 209 complex at a Galactocentric
distance of 21 kpc, one of the outermost sites of star formation
in the Galaxy (Fich \& Blitz 1984).
Molecular outflow activity has been inferred from 
broad CO wings by Wouterloot, Brand \& Henkel (1988).
A H$_2$O maser source has also been detected in the 
vicinity by Cesaroni et al (1988).
The above indicators 
confirm that star formation is still in progress in the S 209 complex.
Despite its large heliocentric distance (12 kpc),
S 209 is expected to be detectable in infrared wavebands due to its 
high intrinsic luminosity.
Surprisingly, no study of the far infrared continuum emission
from the S209 region exists in the literature.

 S 187 is an optical H II region (Sharpless 1959) located
 at the near side of dark cloud L 1317
at a distance of 1 kpc, belonging to the Orion arm in the Galaxy.
High angular resolution radio continuum map of this region
shows the ionized gas to extend over $\sim 6^{\prime}$ with rich structures
(Snell \& Bally, 1986).
Association of this region with a large molecular cloud complex 
has been known since Blair et al (1975) detected extended 
CO emission from this region. 
 Bally \& Lada (1983) found first evidence for high velocity
molecular outflow from S 187, later confirmed to be of extended and
bipolar nature by Casoli, Combes \& Gerin (1984a).
The full extent of this molecular complex has
become more evident from the large scale surveys 
(though with crude gridding), carried out 
in $^{12}$CO and $^{13}$CO lines by Casoli, Combes \& Gerin (1984b) and 
Yonekura et al (1997).
Various evidences of recent
star formation activity in this region have been presented
by Zavagno, Deharveng \& Caplan (1994).
The structurally rich emission from the
molecular as well as the ionized gas, prompted us to study
the emission from the dust component in S 187.

  The next two sections describe the observations and the 
results.

\section{Observations}
\subsection{Balloon-borne observations}

The Galactic star forming regions associated with 
W3(OH), S209 and S187 were
mapped using the 12 channel two band far infrared (FIR)
photometer system at the Cassegrain focus of the  TIFR 100 cm (f/8) balloon 
borne telescope.  
The photometer uses a pair of six element 
(2$\times$3 close packed configuration) composite Silicon 
bolometer arrays cooled to 0.3 K using a closed cycle
$^{3}$He refrigerator and it has been described in Verma et al (1993).
The same region of the sky was viewed simultaneously
in two FIR bands with near identical fields of view of
1$.^{\prime}$6 per bolometer, thus instantaneously
covering an area of 6$.^{\prime}$0 $\times$ 3$.^{\prime}$4 in each band.
The sky was chopped along the cross-elevation axis at 10 Hz 
with a throw of 4$.^{\prime}$2. 
Full details of the 100 cm telescope system and the
observational procedure can be found in Ghosh et al (1988).
These sources were observed in two different balloon flights
with slightly different FIR passbands of the photometer.
The journal of observations and other details are presented in Table 1.
The spectral responses of the two bands, 
relative responses of the detectors, 
absolute calibration of the photometer and other details
specific to these two flights
in 1993 and 1995 have been presented in Ghosh et al (2000) and
Mookerjea et al (1999) respectively.

The observed chopped signals have been 
deconvolved using an indigenously developed procedure based on the
Maximum Entropy Method (MEM) similar to that of Gull
\& Daniell (1978) (see Ghosh et al 1988, for details). 
The accuracy of the absolute aspect of the telescope 
was improved by using a focal plane optical photometer which 
detects stars (in an offset field) while the telescope scans the
FIR target source.
The achieved absolute positional accuracy is $\sim$ 0$.^{\prime}$5.

\subsection{IRAS Data}

  The data from the IRAS survey in the four bands 
(12, 25, 60 and 100 $\mu$m)
for the regions around the three target sources
were HIRES processed (Aumann et al, 1990) at the
Infrared Processing and Analysis Center (IPAC\footnote{IPAC is
funded by NASA as part of the part of the IRAS extended mission
program under contract to JPL.}, Caltech).
These maps have been used for extracting 
sources and quantifying interband positional associations and flux densities.

\section{Results}

\subsection{Intensity Maps}

The MEM deconvolved TIFR maps 
at 148 and 209 $\mu$m and the HIRES processed IRAS maps
at 12, 25, 60 and 100 $\mu$m 
for the Galactic star forming region W3(OH)
has been presented in Figs. 1 and 2 respectively.
Similarly, 
the intensity maps for S209 at 138, 205 $\mu$m and the IRAS bands 
have been presented in a similar format in Figs. 3 and 4. 
Due to limited dynamic range achieved in the 138 $\mu$m band
for S187, the intensity maps for this source are 
shown only at 205 $\mu$m and the IRAS bands
(Fig. 5 \& 6).

Whereas the IRAS maps have very high dynamic ranges ($>$ 1000),
the same for the TIFR maps is restricted to $\sim$ 300 under
the best circumstances. The contour levels displayed in TIFR maps
for each programme source depend 
on the detector noise condition (which varied from time to time)
at the time of the corresponding observations.

The angular resolution achieved in the TIFR bands is 
approximately represented
by the deconvolved sizes of the point-like (planet) source
in respective bands (see Table 1).
All three programme sources, W3(OH), S209 and S187 show
extended emission in both the TIFR bands. 

The angular resolutions in the HIRES processed maps for
each region are listed in Table 2, which depend on the 
observational details like relative 
orientation of scan tracks of the telescope boresight
among different HCONs (Aumann, Fowler \& Melnyk 1990).
Although extensions are seen in many IRAS bands, the TIFR maps
show superior angular resolution as a result of their smaller
and circular beam.

 Discrete sources have been extracted from the TIFR and HIRES maps
using a procedure described in Ghosh et al (2000).
The longest wavelength channel (TIFR Ch-II) map has been used
as the primary band. The sources detected in this 
are associated with sources in other bands if they 
satisfy the positional match criterion ($<1^{\prime}$ separation 
with TIFR Ch-I and $<1.^{\prime}$5 for HIRES bands).
 A total of nine sources in all three regions have been detected, 
details of which are listed in Table 3. Six of these have been
detected in both the TIFR bands.
All these nine sources have an association with HIRES source in 
at least one band (8 have associations in 2 or more IRAS bands).
The listed flux densities have been obtained by 
integrating over a circle of 3$^{\prime}$ ~diameter.
Six of these
also appear in the IRAS Point Source Catalog (hereafter PSC).
The PSC flux densities are also listed for comparison with those
obtained from the HIRES maps.
Four of these six PSC sources have upper limits in at least
one IRAS band. This reflects the complexiety of the morphology of
these regions.
The dust temperatures in the FIR, T${}_{FIR}$,
have been computed from the flux densities in the TIFR bands,
assuming an emissivity
law of $\epsilon_{\lambda} \propto {\lambda}^{-2}$. These are also
listed in Table 3.

\subsubsection{W3(OH)}

 Strong emission is seen in both 148 and 209 $\mu$m bands
from W3(OH) and the peak position (S2) matches with that of the
IRAS PSC source 02232+6138 (Fig. 1). The corresponding source is also
the strongest in all the four HIRES maps (Fig. 2). 
Whereas in TIFR bands S2 is resolved, it is pointlike in the IRAS
bands. There are two other sources detected
in both the TIFR maps. The second brightest source (S3) has counterparts
in 12, 25 and 60 $\mu$m maps and a clear extension in 100 $\mu$m map.
The diffuse emission has been detected in all the six bands.

 The extension of the isophot contours towards NE of W3(OH)
in the TIFR bands, match remarkably well with the plume 
($\sim 2 pc \times 1.3 pc$)
seen in the recent mapping in NH$_3$ line by Tieftrunk et al (1998).
In fact they concluded that the W3(OH) core is much larger
than thought earlier. 

 Combining TIFR data along with the sub-millimeter 
measurement of Chini et al (1986), the dust emissivity index
is found to be 1.8 between 200 and 350 $\mu$m. 

  The total emission from 
a circular region of 16$^{\prime}$ diameter around the strong peak W3(OH), 
are 9601 and 6305 Jy at 148 and 209 $\mu$m respectively.
The fraction of this in diffuse emission has been estimated to
be 15\% and 13\% respectively by subtracting the contributions
from the detected discrete sources (Table 3).
A similar analysis of the IRAS-HIRES maps of the same region
has quantified the diffuse emission to be 85\%, 55\%, 56\% and
71\% at 12, 25, 60 and 100 $\mu$m respectively.
It may be noted that since the mapping in TIFR bands are carried
out in sky chopped mode (in contrast to the IRAS bands), some
part of the diffuse emission with low spatial gradient 
could have been missed in these bands.
The total infrared luminosty estimated from the entire region 
is 1.91$\times 10^{5}$ $L_{\odot}$.

\subsubsection{S 209}

 There is a good correlation and structural similarity
between the extended emission from
the dust component in all the six bands.
The complex emission structure has restricted the reliable
source extraction for the IRAS Point Source Catalog as evident
from inconsistent flux densities in different bands for the
main source corresponding to S 209, IRAS 04073+5102.
However, numerical aperture photometry on HIRES processed 
IRAS maps provide reliable estimates of flux densities.

 The main source in S209 region is clearly resolved into 
two sources (S5 \& S6) at 205 $\mu$m map and there is indication for 
the same in the 138 $\mu$m map (Fig. 3). The strongest peaks in both the
TIFR bands (S5) coincide with the position of IRAS 04073+5102.
The neighbouring source S6 is associated with IRAS 04072+5100.
In 25 and 60 $\mu$m IRAS bands, S6 is clearly seen and an indication
is present in the 12 $\mu$m map (Fig. 4). The 100 $\mu$m HIRES processed
map does not resolve S5/S6.

 Chini et al (1984) have detected S 209 in 1-mm continuum and presented
the thermal emission from the dust after correcting for the
expected free free emission from the hot gas. Using the 
flux densities at 205 $\mu$m and 1-mm, a very flat dust emissivity
exponent of 0.64 has been found for this sub-mm region.
In case the emission at 1-mm originates from a different colder dust
component, then the above index is an underestimate.

  The total emission from the S 209 region presented in Fig. 3
is 5548 and 4000 Jy at 138 and 205 $\mu$m respectively.
The fraction of this in diffuse emission has been estimated to
be 57\% and 46\% respectively by subtracting the contributions
from the detected discrete sources.
A similar analysis of the IRAS-HIRES maps of the same region
(Fig. 4)
has quantified the diffuse emission to be 52\%, 44\%, 49\% and
77\% at 12, 25, 60 and 100 $\mu$m respectively.
Hence right through the mid and far infrared region, a good
part of the emission is in diffuse form.
The total infrared luminosty estimated from the entire region 
is 2.0$\times 10^{6}$ $L_{\odot}$ (for distance = 12 kpc).

 Balser et al (1995) have modelled their 8.7 GHz radio continuum
measurements of the S 209 region ($8^{\prime}{\times}8^{\prime}$) 
and they conclude 
that the exciting source is either a ZAMS O6.5 star or a O5
star depending on the data used (VLA / MPIR). These stellar
types correspond to a luminosity of 1.5 x 10$^5$ $L_{\odot}$
or 6.8 x 10$^5$ $L_{\odot}$ respectively (Thompson 1984).

\subsubsection{S 187}

 The dynamic range of the TIFR maps of S 187 region is rather
limited due to larger than usual noise in the bolometer channels
during these observations (only the 205 $\mu$m map presented
here). The strongest source (S9) associated with
IRAS 01202+6133 is resolved at 205 $\mu$m (Fig. 5).
The morphology of the diffuse emission in the TIFR band
resembles the same in the HIRES maps (Fig. 6).
This is despite the fact that TIFR observations used sky chopping
whereas IRAS survey did not.
The emissions in all the five bands are dominated by the source 
associated with IRAS 01202+6133. Most of the additional
emission originates from an annular ring like structure of diamater
$\sim 10^{\prime}$. The ionized gas resides at the central
cavity of the annular region as inferred from high resolution
radio continuum map at 1.4 GHz (Snell \& Bally, 1986).
The position of the high velocity molecular outflow lies about
$2^{\prime}$ west of S9. The H$_{2}$O maser source detected by Henkel,
Haschick \& Gusten
(1986) is positionally very close to the outflow source.
No local enhancement can been observed in any of the TIFR or HIRES 
bands at the location of the H$_{2}$O maser / outflow source.
The position of the NH$_3$ core detected in the S 187 region
(Jijina, Myers \& Adams, 1999),
also does not show positional match with any peak in the maps
of dust continuum emission.

  The total emission from the S 187 region presented in Fig. 5
is  7256 Jy at 205 $\mu$m.
The fraction of this in diffuse emission has been estimated to
be 55\% by subtracting the contributions
from the detected discrete sources.
A similar analysis of the IRAS-HIRES maps of the same region
(Fig. 6)
has quantified the diffuse emission to be 66\%, 68\%, 53\% and
69\% at 12, 25, 60 and 100 $\mu$m respectively.
For this source too, right through the mid and far infrared region, a large
part of the emission is in diffuse form, which is quite expected
considering the complex morphology of the region.
The total infrared luminosty estimated from the entire region 
is 1.7$\times 10^{4}$ $L_{\odot}$ (for distance = 1 kpc).

 Using the mass of the molecular cloud associated with
S 187, as estimated by Yonekura et al (1997) from their $^{13}$CO survey
(their cloud \# 164),
we determine the average luminosity per unit mass to be
$\sim$ 2.2 $L_{\odot}/M_{\odot}$. This value is very similar to
that found for W 31 star forming complex (Ghosh et al 1989).

\subsection{Distribution of dust temperature and optical depth}

Taking advantage of the nearly identical circular beams of
the TIFR bands and the
simultaneity of observations, reliable 
maps of dust temperature
and optical depth (at 200 $\mu$m, $\tau_{200}$) have 
been generated for W3(OH) and S209 regions.
The available dynamic ranges in both the TIFR bands for
these two sources allow us to meaningfully determine the
dust temperature and optical depth distributions.
These are presented in Figs. 7 and 8 respectively.
A dust emissivity law of
$\epsilon_{\lambda}\propto\lambda^{-2}$ has been assumed for
this purpose. Details of the procedure can be found in
Ghosh et al (2000). 

  For W3(OH) region, the distribution of $\tau_{200}$ 
shows a peak near the intensity peak but the T(148/209)
distribution shows a plateau there (Fig. 7). Some regions of 
higher dust temperature are also seen.
The second peak in the optical depth map clearly corresponds to 
the matter distribtuion around 02236+6142.
The dust temperature at the position of IRAS 02232+6138
is 21 K, whereas the kinetic temperature
of the NH$_3$ component has been found to be 27 K (Tieftrunk et al 1998).
In addition, the shape of the $\tau_{200}$ distribution
around IRAS 02232+6138 (within $2^{\prime}$) resembles the gas distribution
traced by NH$_3$.
A detailed comparison 
should help understanding the gas-dust coupling
in denser regions of star formation, like the 
interstellar environment around W3(OH).

 The T(138/205) and $\tau_{200}$ distributions for S209
show an almost anticorrelation (Fig. 8). The hotspots are 
located near the two resolved sources in the 205 $\mu$m map.
Most of the region in S 209 has the dust temeparture higher than
27 K and the presence of colder dust is limited to the
outermost perifary. This is in contrast to W3(OH) region
where the dust is relatively cooler.

\vskip 0.5cm
\centerline{\bf Acknowledgements}
\vskip 0.5cm

We thank 
S.L. D'Costa, M.V. Naik, S.V. Gollapudi, D.M. Patkar, 
M.B. Naik and G.S. Meshram for their support for the
experiment. The members of TIFR Balloon Facility (Balloon group and
Control \& Instrumentation group), Hyderabad, are thanked for their roles
in conducting the balloon flights. IPAC is thanked for providing HIRES
processed IRAS data. 
SKG thanks the Institute of Space and Astronautical Science (ISAS), Japan,
for their hospitality, where part of the work was done.

\vskip 0.5cm

%%%%%%%%%%%%%%%%%%%%%%%%%%%% End of Main Text %%%%%%%%%%%%%%%%%%%%%%%%%%%%%%
%%%%%%%%%%%%%%%%%%%%%%%%%%%% Bibliography begins %%%%%%%%%%%%%%%%%%%%%%%%%%%

%%%%%%%%%%%%%%%%%%%%%%%%%%%% Bibliography ends %%%%%%%%%%%%%%%%%%%%%%%%%%%%%
  
\vfill \eject
\clearpage
\newpage

%%%%%%%%%%%%%%%%%%%%%%% Fig1 begins
\begin {figure*}[p]
\vskip 5cm
\epsfxsize=350.0pt
\epsfbox{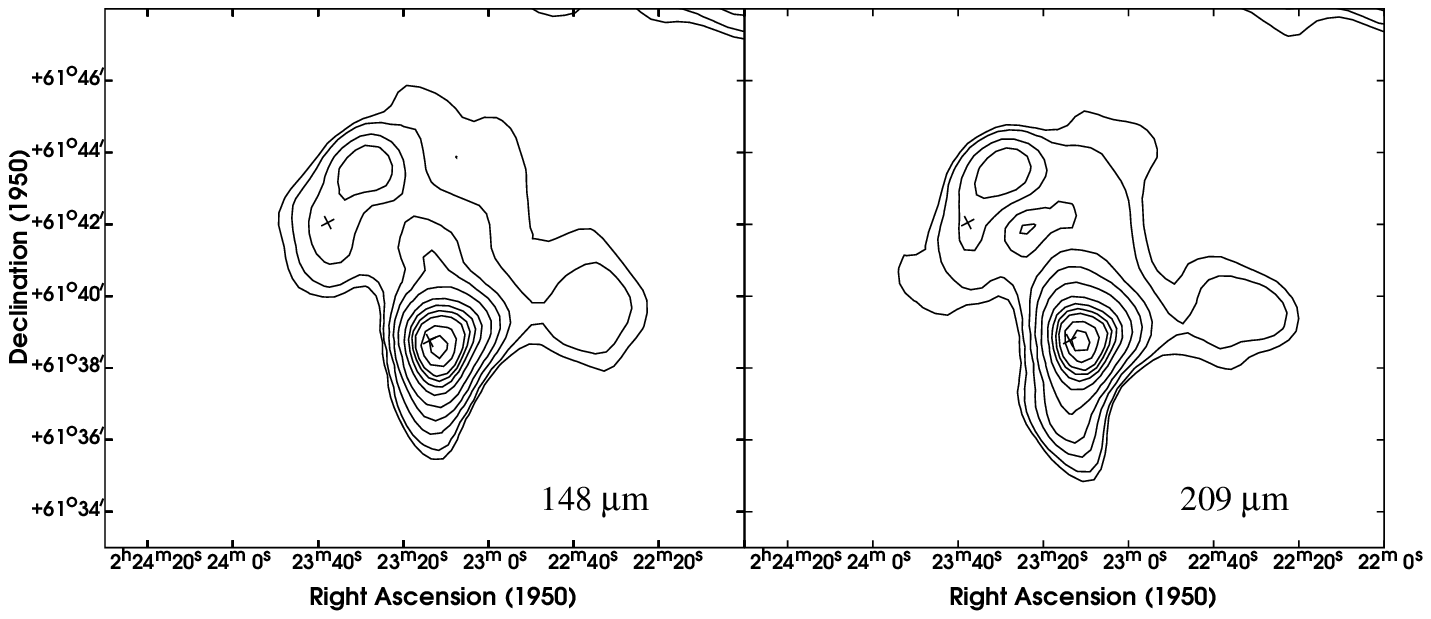}
\caption{
The intensity maps for the region around W3(OH) in TIFR
bands --
{\bf (a)} at 148 $\mu$m with peak = 3059 Jy/sq. arcmin,
{\bf (b)} at 209 $\mu$m  with peak = 2101 Jy/sq. arcmin.
The isophot contour
levels in both (a) and (b) are 90, 70, 50, 40, 30, 20, 10, 5, 
2.5, 1 \& .5 \% of the respective peaks. 
The crosses denote the positions of the IRAS PSC sources
02232+6138 (main source) \& 02236+6142.
}
\end {figure*}
%%%%%%%%%%%%%%%%%%%%%%% Fig1 Ends
%\clearpage
%\newpage
%%%%%%%%%%%%%%%%%%%%%%% Fig2 begins
\begin {figure*}[p]
\vskip 3cm
\epsfxsize=350.0pt
\epsfbox{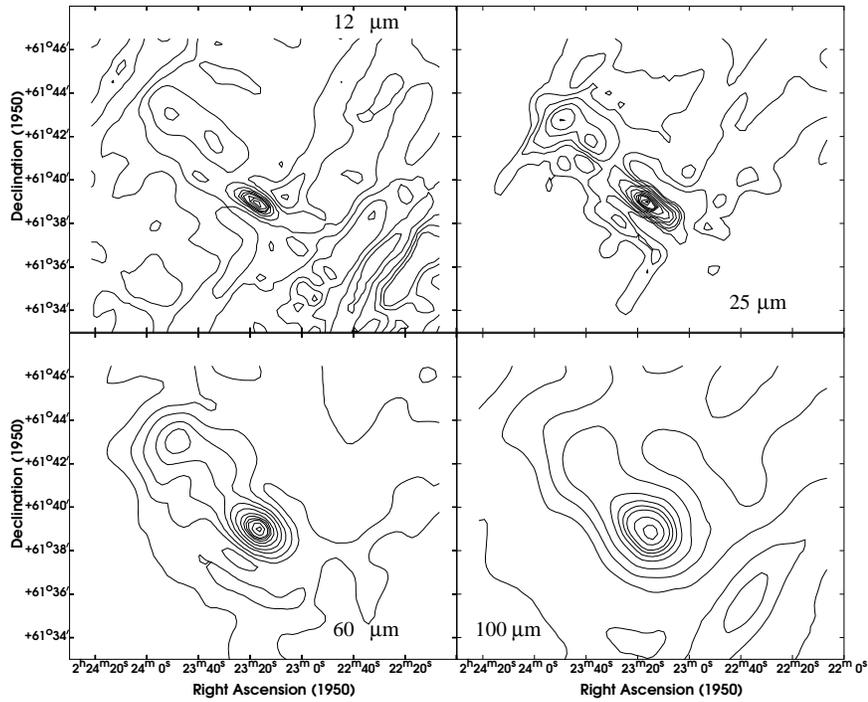}
\caption{
The HIRES processed IRAS maps for a similar region 
around W3(OH), as shown in Fig. 1,
in the four bands --
{\bf (a)} at 12 $\mu$m with peak = 66.9 Jy/sq. arcmin,
{\bf (b)} at 25 $\mu$m  with peak = 1420 Jy/sq. arcmin.
{\bf (c)} at 60 $\mu$m  with peak = 9870 Jy/sq. arcmin.
{\bf (d)} at 100 $\mu$m  with peak = 3140 Jy/sq. arcmin.
The isophot contour levels in (a) (b) \& (c) are 
90, 70, 50, 40, 30, 20, 10, 5, 2.5, 1, .5, \& .25 \% 
of the respective peaks.
In (d), only the higher 11 of these contours have been displayed.
}
\end {figure*}
%%%%%%%%%%%%%%%%%%%%%%% Fig2 Ends

%\clearpage
%\newpage
%%%%%%%%%%%%%%%%%%%%%%% Fig3 begins
\begin{figure*}[p]
\vskip 3cm
\epsfxsize=350.0pt
\epsfbox{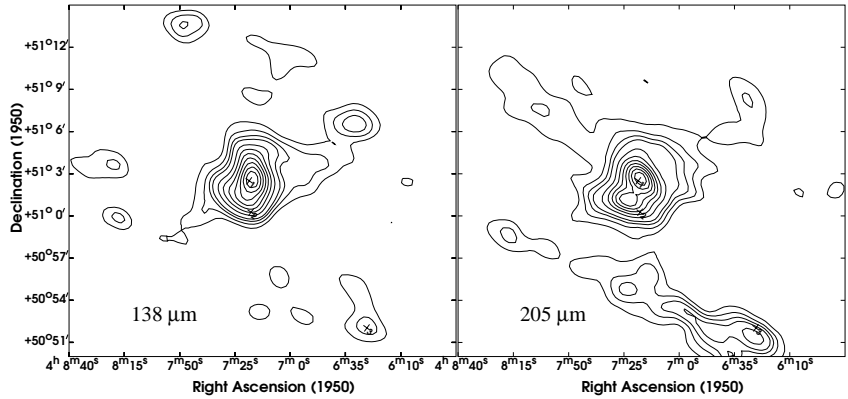}
\caption{
The intensity maps for the region around S209 in TIFR
bands --
{\bf (a)} at 138 $\mu$m with peak = 441 Jy/sq. arcmin,
{\bf (b)} at 205 $\mu$m  with peak = 213 Jy/sq. arcmin.
The isophot contour
levels in both (a) and (b) are 
90, 80, 70, 60, 50, 40, 30, 20, 15, 10, 
\& 5 \% of the respective peaks. 
The crosses denote the positions of the IRAS PSC sources
04073+5102 (main source), 04072+5100 (the nearby source
which is resolved in 205 $\mu$m map) and 04064+5052.
}
\end{figure*}
%%%%%%%%%%%%%%%%%%%%%%% Fig3 Ends
%\clearpage
%\newpage

%%%%%%%%%%%%%%%%%%%%%%% Fig4 begins
\begin {figure*}[p]
\vskip 3cm
\epsfxsize=350.0pt
\epsfbox{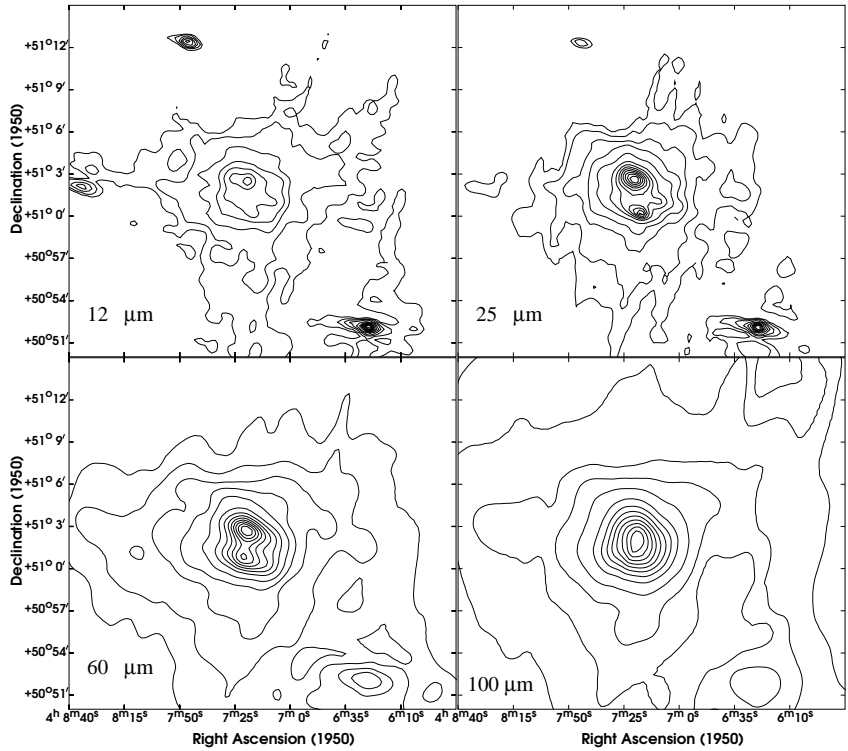}
\caption{
The HIRES processed IRAS maps for
a similar region around S209, as shown in Fig. 3,
in the four bands --
{\bf (a)} at 12 $\mu$m with peak = 374 Jy/sq. arcmin,
{\bf (b)} at 25 $\mu$m  with peak = 5470 Jy/sq. arcmin.
{\bf (c)} at 60 $\mu$m  with peak = 12300 Jy/sq. arcmin.
{\bf (d)} at 100 $\mu$m  with peak = 5830 Jy/sq. arcmin.
The isophot contour levels in (a) are 
30, 20, 10, 5, 2.5, 1 \& .5 \% of the peak, and in
all the other three bands are 
90, 80, 70, 60, 50 ,40, 30, 20, 10, 5, 2.5, 1 \& .5 \% 
of the respective peaks.
}
\end {figure*}
%%%%%%%%%%%%%%%%%%%%%%% Fig4 Ends

%\clearpage
%\newpage

%%%%%%%%%%%%%%%%%%%%%%% Fig5 begins
\begin{figure*}[p]
\vskip 2cm
\epsfxsize=300.0pt
\epsfbox{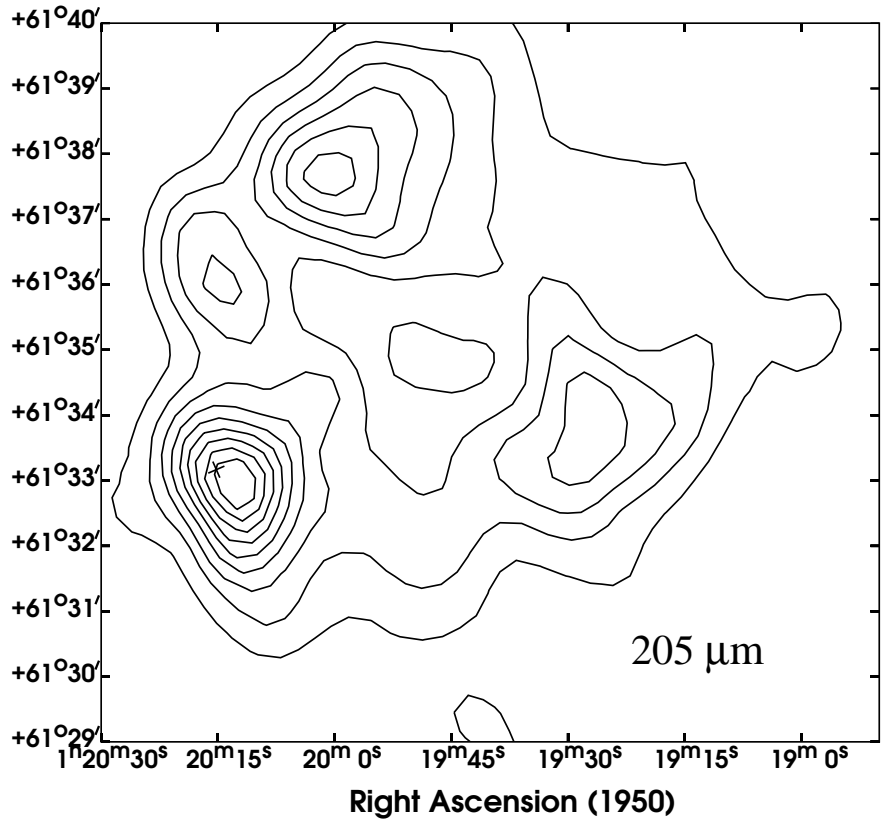}
\caption{
The intensity map for the region around S187 in TIFR
band at 205 $\mu$m.
The isophot contour levels 
are 90, 80, 70, 60, 50, 40, 30, 20 \& 10 \% of the 
peak intensity (388 Jy/sq. arcmin).
The cross denotes the position of the IRAS PSC source 01202+6133
(main source).
}
\end{figure*}
%%%%%%%%%%%%%%%%%%%%%%% Fig5 Ends
%\clearpage
%\newpage

%%%%%%%%%%%%%%%%%%%%%%% Fig6 begins
\begin {figure*}[p]
\vskip 3cm
\epsfxsize=350.0pt
\epsfbox{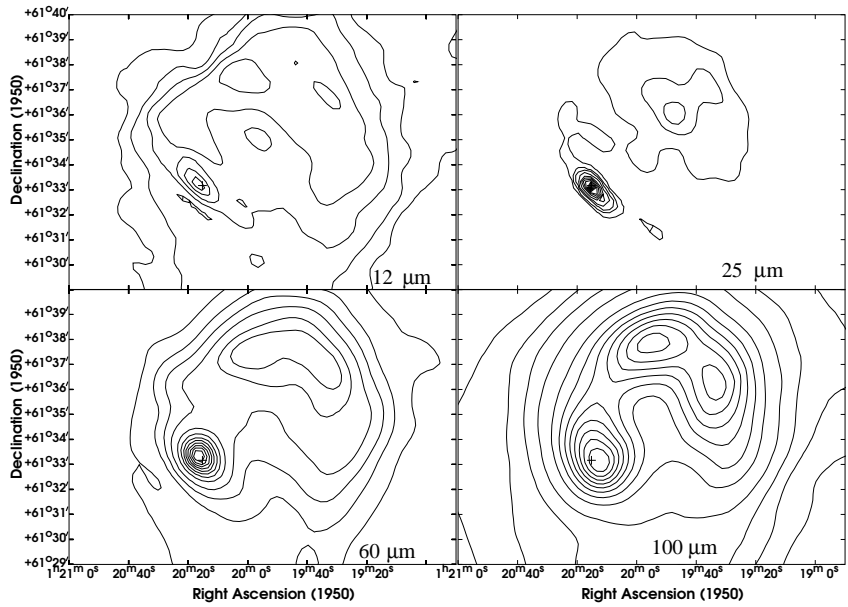}
\caption{
The HIRES processed IRAS maps for a similar region 
around S187, as shown in Fig. 5,
in the four bands --
{\bf (a)} at 12 $\mu$m with peak = 48.8 Jy/sq. arcmin,
{\bf (b)} at 25 $\mu$m  with peak = 557 Jy/sq. arcmin.
{\bf (c)} at 60 $\mu$m  with peak = 754 Jy/sq. arcmin.
{\bf (d)} at 100 $\mu$m  with peak = 460 Jy/sq. arcmin.
The isophot contour levels in (a) are 
30, 20, 10, 5, 2.5 \& 1 \% of the peak (the peak is outside the 
region displayed here)
and in (b), (c) \& (d) are 
90, 80, 70, 60, 50, 40, 30, 20, 10, 5, 2.5, \& 1 \% 
of the respective peaks.
}
\end {figure*}
%%%%%%%%%%%%%%%%%%%%%%% Fig6 Ends
%\clearpage
%\newpage

%%%%%%%%%%%%%%%%%%%%%%% Fig7 begins
\begin{figure*}[p]
\vskip 3cm
\epsfxsize=350.0pt
\epsfbox{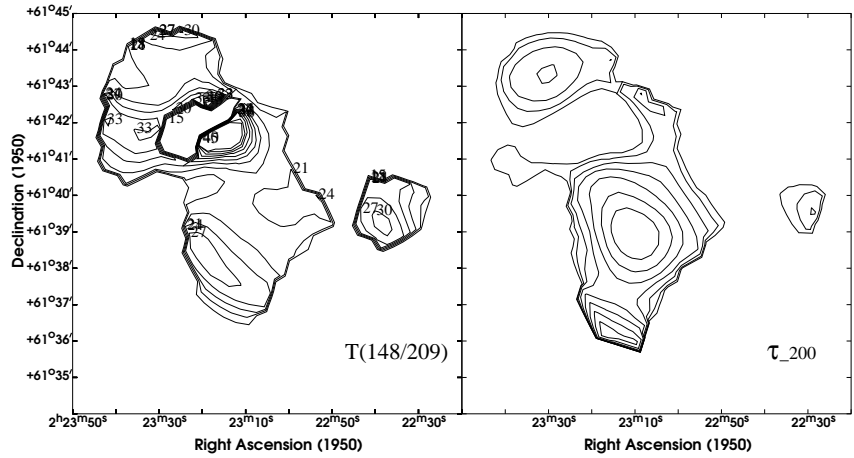}
\caption{
The distribution of dust temperature T(148/209), and optical depth at 
200 $\mu$m, $\tau_{200}$, for the region around
W3(OH) assuming a dust emissivity law of 
$\epsilon_{\lambda} \propto \lambda^{-2}$.
The isotherms correspond to 15 K to 36 K in steps of 3 K, 40 \& 45 K.
Temperature values are displayed near the contours.
The highest contour of $\tau_{200}$ (innermost at the bottom)
corresponds to a value of 0.16 and the successive contours
represent values reducing by factor of 2.
}
\end{figure*}
%%%%%%%%%%%%%%%%%%%%%%% Fig7 Ends
%\clearpage
%\newpage

%%%%%%%%%%%%%%%%%%%%%%% Fig8 begins
\begin{figure*}[p]
\vskip 3cm
\epsfxsize=350.0pt
\epsfbox{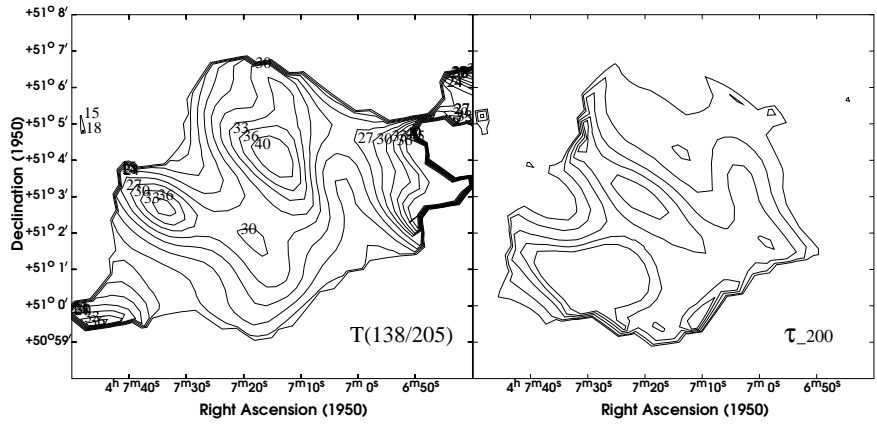}
\caption{
The distribution of dust temperature T(138/205), and optical depth at 
200 $\mu$m, $\tau_{200}$, for the region around
S209 assuming a dust emissivity law of 
$\epsilon_{\lambda} \propto \lambda^{-2}$.
The isotherms refer to the same temperatures as in Fig. 7.
The $\tau_{200}$ contours represent 100, 75, 50, 25 \& 12.5 \% of the 
peak value of 0.67.
}
\end{figure*}
%%%%%%%%%%%%%%%%%%%%%%% Fig8 Ends

%\vfill
\clearpage
%\newpage
%-- Table 1--------------------- Two flights 
%
\begin{center}
\begin{table}
 \vskip 7cm
\caption{The journal and other observational details}
\vskip 0.5cm
\begin{tabular}{|c|c|c|c|c|c|c|}
\hline
Flight date & FIR & $\lambda_{eff}$ & $\lambda_{eff}$ &
Planet & Planet   & Planet \\
  & target & Ch-I & Ch-II  & used & 
FWHM & FWHM  \\
&& ($\mu$m) & ($\mu$m) && Ch-I & Ch-II \\
\hline
18-Nov-1993 & W3(OH) & 148 & 209 & Jupiter &
$1.^{\prime}0 \times 1.^{\prime}4$  & 
$1.^{\prime}0 \times 1.^{\prime}3$  \\
\hline
12-Nov-1995 & S209 & 138 & 205 & Saturn &
$1.^{\prime}6 \times 1.^{\prime}9$  & 
$1.^{\prime}6 \times 1.^{\prime}8$  \\
& S187 &&&&&\\
\hline
\end{tabular}
\end{table}
\end{center}

%---------------------------------- End Table 1
%\vfill
 \clearpage
%\newpage

%-- Table 2---------------------HIRES resolutions 
%
\begin{center}
\begin{table}
 \vskip 9cm
\caption{Angular resolutions in the HIRES maps}
\vskip 0.5cm
\begin{tabular}{|ccccc|}
\hline
Source & resolution & resolution & resolution & resolution \\
region  & at 12 $\mu$m & at 25 $\mu$m & at 60 $\mu$m & at 100 $\mu$m \\
 & FWHM & FWHM  & FWHM  & FWHM  \\
\hline
W3(OH) & $74'' \times 29''$ & $66'' \times 31''$ & $114'' \times 57''$ & 
$74'' \times 29''$ \\
\hline
S209 & $56'' \times 27''$ & $54'' \times 28''$ & $91'' \times 54''$ & 
$110'' \times 64''$ \\
\hline
S187 & $39'' \times 27''$ & $41'' \times 28''$ & $76'' \times 46''$ & 
$113'' \times 97''$ \\
\hline
\end{tabular}
\end{table}
\end{center}

%---------------------------------- End Table 2
%\vfill
%\clearpage
%\newpage

%---------------- Table 3 begins --------------------
\hoffset -2.5cm
\voffset -2.5cm
\begin{center}
\begin{table*}
\caption{Position and flux density details of the detected sources.}
\vskip 0.5cm
\begin{tabular}{ccccrrrrrrc}
\hline
\#& RA & Dec & IRAS PSC&\multicolumn{6}{c}{Flux Density (Jy)} & T${}_{FIR}^{b}$ \\
\cline{5-10}
& (1950)    &(1950)  &  associations & \multicolumn{2}{c}{TIFR} & 
\multicolumn{4}{c}{IRAS$^a$} &\\
& h ~~m ~~s & $^o$ ~~$'$ ~~$''$ &&    209 / & 148 / &   
100 $\mu$m& 60 $\mu$m& 25 $\mu$m& 12 $\mu$m & (K) \\
&&&& 205 $\mu$m & 138 $\mu$m &&&&& \\
\hline
&{\it W3(OH)} & {\it region} &&&&&&&&\\
S1& 2 22 28.3 & +61 39 32& ... & 128 &260 & -- & -- & 35 
& 30 & 30 \\
S2& 2 23 11.3 & +61 38 54& 02232+6138 & 4872& 7198&  10030 &  
9488 & 683&  66 & 22 \\
&"&"&"&&&     10600$^c$ & 9269$^c$ &   535$^c$ & 40.6$^c$ &\\
S3& 2 23 29.0 & +61 43 34& 02236+6142 & 476 & 744 &  1886 &  
901 & 67 & 36 & 23 \\
&"&"&"&&&   $<$10600$^c$ & $<$1712$^c$ &   $<$91$^c$ & 12.3$^c$ &\\
&&&&& \\
&{\it S209} & {\it region} &&&&&&&&\\
S4& 4 06 26.7 & +50 51 18& 04064+5052 & 323& 224&  190 &  
149 & 48&  30 & 15 \\
&"&"&"&&&     171.2$^c$ & 83.4$^c$ &  43.3$^c$ & 25.4$^c$ &\\
S5& 4 07 17.4 & +51 02 45& 04073+5102 & 970& 2139&  2298 &  
2159 & 409 & 70 & 31 \\
&"&"&"&&&     2723$^c$ & $<$0.4$^c$ & 97$^c$ & 16$^c$ &\\
S6& 4 07 22.6 & +51 01 07& 04072+5100 & 873& -- & -- &  
1905 & 236 & 38 & -- \\
&"&"&"&&&     $<$2700$^c$ & $<$439$^c$ & 41$^c$ & 2.4$^c$ &\\
&&&&& \\
&{\it S187} & {\it region} &&&&&&&&\\
S7& 1 19 25.7 & +61 34 02& ... & 885 & 1770 & -- & 829 & --
& 41 & 28 \\
S8& 1 20 01.3 & +61 37 45& ... & 1094 & -- & 2005 & 1310 & --
& 52 & -- \\
S9& 1 20 13.6 & +61 32 52& 01202+6133 & 1304& -- &  1908 &  
1400 & 211 & 32 & -- \\
&"&"&"&&&     $<$1700$^c$ & 882$^c$ & 182$^c$ & 10.4$^c$ &\\
&&&&& \\
\hline
\end{tabular}
\vskip 0.3cm
${}^a$ From HIRES processed maps unless specified otherwise. The 
flux densities are integrated over a circle of $3.'0$ diameter. \\
${}^b$ Determined using the flux densities in TIFR bands and assuming a
gray body spectrum with emissivity $\epsilon_{\lambda} \propto {\lambda}^{-2}$. \\
${}^c$ From IRAS Point Source Catalog. \\
\end{table*}
\end{center}
\vfill \eject
%---------------- Table 3 Ends --------------------

\end{document}